\documentclass[tightenlines,preprint,amsmath,amssymb,aps,showpacs,superscriptaddress]{revtex4-1}
\usepackage{mathptmx}
\usepackage{helvet}
\usepackage[latin9]{inputenc}
\usepackage{color}
\usepackage{amsmath}
\usepackage{amssymb}
\usepackage[unicode=true,pdfusetitle,
 bookmarks=true,bookmarksnumbered=false,bookmarksopen=false,
 breaklinks=false,pdfborder={0 0 0},pdfborderstyle={},backref=false,colorlinks=true]
 {hyperref}
\usepackage{breakurl}

\makeatletter
\numberwithin{equation}{section}

\makeatother

\begin{document}

\title{Torsion driven Inflationary Magnetogenesis}

\author{Rahul Kothari}
\affiliation{Department of Physics and Astronomy, The University of the
Western Cape, Bellville 7535, South Africa}
\author{M.V.S. Saketh}
\affiliation{Department of Physics, University of Maryland, College Park 
MD 20742-4111, USA}
\author{Pankaj Jain}
\affiliation{Department of Physics, Indian Institute of Technology, Kanpur 208016, India}

\begin{abstract}
We show that breaking of the conformal invariance of electromagnetic Lagrangian which is required for inflationary magnetogenesis arises
naturally in the
Poincar{\'e} Gauge Theory. We use the minimal coupling prescription to introduce the electromagnetic gauge fields as well as non-abelian gauge fields in this theory. Due to the addition of non-abelian gauge fields, we show that the solar constraints on this model can be naturally evaded.
We find that in the minimal version of this model the 
generated magnetic field is too small to explain the observations. We discuss some generalizations of the gravitational action, including the Starobinsky model and a model with conformal invariance. We show that such generalizations naturally generate the kinetic energy terms required for magnetogenesis. 
We propose a generalization of the minimal model by adding a potential term, which is allowed within the framework of this model, and show that it leads to
sufficiently large magnetic fields. 
\end{abstract}
\maketitle

\section{Introduction}

Magnetic fields are found in almost all bound structures, such as, galaxies, 
stars, star clusters, and even some planets like Earth
\citep{Widrow:2011hs,Subramanian:2015lua,Kerstin2013}. 
Recent observational evidences
\citep{Neronov2010,Tavecchio2010,Tavecchio2011,Caprini2015} suggest
that even cosmic voids contain magnetic fields of order $10^{-16}$ Gauss with correlation lengths
of 1 Mpc or more. The upper limit on the magnetic field on such scales is of order $10^{-9}$ Gauss 
\cite{PhysRevD.82.083005}.
Such large scale correlation is hard to explain by
any astrophysical process. This is because one would expect astrophysical
processes to generate these fields during radiation domination phase
and small Hubble radius during this epoch cannot account for such
large scale correlations.
Even large scale magnetic fields in galaxies \cite{Ferriere2009} 
 and clusters of galaxies
in the distant past are not easy to explain. 
 One possibility is that these fields are
actually a relic from the inflationary era \citep{Turner:1987bw,Kandus:2010nw,Durrer:2013pga}. That
explains why they are supposedly present in the voids. The same mechanism
also explains the origin of strong magnetic field in bound structures
by serving as the seed field that was amplified via dynamo mechanism
\citep{Moffatt1976} in these structures.

There exist many models of inflationary magnetogenesis in which the 
 magnetic field is amplified many folds during inflation 
 \citep{Turner:1987bw,Ratra1992,Dolgov:1993vg,Gasperini:1995dh,Giovannini:2000ad,Martin:2007ue,Subramanian:2009fu,Atmjeet:2013yta,Fujita:2015iga,Nandi:2017ajk}. All these model break conformal invariance of electromagnetic Lagrangian
which is required to amplify
 the vacuum fluctuations of electromagnetic
fields. In the model
 developed by Ratra \citep{Ratra1992}, the electromagnetic
Lagrangian is coupled with the inflaton field to break the conformal
invariance. It turns out that for some parameter values this model suffers 
from strong coupling
or the back reaction problem \cite{Demozzi:2009fu,Fujita:2012rb}. 
There exist several proposals 
 \citep{Ferreira2013,Tsagas:2014wba,Campanelli2015} 
to generate cosmic magnetic fields of the
required strength without facing either of these problems, if conformal
invariance in broken during inflation.
However there also exist additional constraints from CMB which impose further
limits on the magnetic field that can be generated \cite{Jain:2012vm,Ferreira:2014hma,Seshadri:2009sy}.

 Although there exist 
several proposals \citep{Turner:1987bw,Dolgov:1993vg,Gasperini:1995dh,Giovannini:2000ad,Martin:2007ue,Subramanian:2009fu,Atmjeet:2013yta,Fujita:2015iga,Nandi:2017ajk}
which break conformal invariance, these do not  
 explain the coupling between electromagnetic and
scalar field assumed in the Ratra model \cite{Ratra1992}. 
So far this coupling has been put in by hand. 
The only
exception is the string inspired model \cite{Gasperini:1995dh} which has some
similarity to the Ratra model.
 We propose that such coupling naturally arises in Poincar{\'e}
Gauge Theory \cite{Kibble1961}, when torsion, a dynamical variable of the theory is
taken into account. It turns out that by a specific choice of torsion,
sourced by a scalar field $\phi$, referred to as ``tlaplon'' in the literature
\citep{Hojman1978}, one can create a scenario where both inflation
and magnetogenesis is driven by the tlaplon itself.
In this formalism a restricted version of torsion is coupled directly 
to the electromagnetic field in a gauge invariant manner.
It has been argued in the literature 
\cite{Kibble1961,1973GReGr...4..333H,hehl1976} 
that an alternative procedure is
to not couple torsion to photons.  
As mentioned in \cite{Kibble1961} this has the strange consequence that for
photon field the spin tensor vanishes whereas this would not be the case 
for a massive vector field. This has been justified by the argument
that photon is special since it is massless 
\cite{1973GReGr...4..333H,hehl1976}. 
However this  
 may lead to inconsistencies.
 A particular problem that we discuss in this paper is the implications
of this procedure if the mass of the vector field is
generated by the Higgs mechanism. In this case we would not couple the
gauge field to torsion. However the gauge field becomes massive due to 
Higgs mechanism and this massive field would now not couple to torsion either.
This is inconsistent since a massive field with spin must couple to torsion.

There also exists considerable literature 
\cite{Muzinich:1984yj,German:1985tj,Katanaev:1985si,Oh:1989wv,German:1993bq,Zhang:1996ee,Kohler:1998ah}
which attempts to derive the coupling
of the gauge fields to torsion by dimensional reduction of a 5-dimensional
theory using the Kaluza-Klien formalism. This has the advantage of preserving
both Poincar{\'e} gauge invariance and electromagnetic gauge invariance in 
4-d and hence can settle the question of how electromagnetic field should 
be introduced in such a theory. 
In particular, it is argued in \cite{German:1985tj} that under certain
conditions the formalism proposed in  \citep{Hojman1978} can be derived
from the 5-dimensional theory. Essentially \cite{German:1985tj} assume
a limited form of torsion in 5-dimension and expresses it in terms of a
scalar field $\Omega$. This is analogous to the procedure used  
in \citep{Hojman1978} in 4-dimensions. Furthermore it is 
shown that the field $\Omega$, and hence the torsion potential $\phi$, 
is related to the $g_{55}$ component of the metric. 
 However this was shown to be invalid in
\cite{Oh:1989wv,German:1993bq} and hence this formalism does
not provide a guidance of how to couple electromagnetic field to torsion.
A more detailed analysis of the 5-dimensional theory 
is provided in \cite{Zhang:1996ee}. In
this paper the authors do not use any specific form of torsion
and use the anholonomic horizontal lift basis for reduction to 4 dimensions. 
The authors find that the electromagnetic field decouples from
torsion. This is easily understood. In \cite{Zhang:1996ee} the authors use only one term in the gravitational action, i.e. the five dimensional analogue of the Ricci scalar $\hat R$, including the contribution due to torsion (Eq. 2.6 of \cite{Zhang:1996ee}). This term does not generate any derivatives of the torsion field. Hence starting from 5 dimension action we do not expect any terms involving derivatives of the electromagnetic field $A_\mu$ in the terms associated with torsion. Since a term independent of derivatives of $A_\mu$ cannot be gauge invariant, we do not expect such terms in the action. Although this is the standard form of the action, other terms, such as $\hat R^2$, can also included \cite{Hayashi1979}. With these terms we do, in general, find a coupling between torsion and electromagnetic field (Jain and Kothari, work in progress). Although this does not provide a 
derivation of the coupling proposed in \citep{Hojman1978}, it does
 suggest that a coupling should exist. 
 The Hojman et al \citep{Hojman1978} 
 procedure may be regarded as an effective simple framework to implement
such a coupling.

The article is structured in the following manner. In Section \eqref{sec:A-Brief-Review},
we discuss in brief the possibility of primordial origin of magnetic
fields, the 
Ratra's model \citep{Ratra1992} and its drawbacks. This is followed
by a brief introduction to the Poincar{\'e} Gauge Theory of Gravity in
Section \eqref{sec:Minimal-coupling-and}. It is found that generically
coupled torsion (a dynamical variable in Poincar{\'e} Gauge Theory) to
electromagnetic field breaks gauge invariance. But this can be avoided
by choosing a restricted form of torsion, as has been suggested in 
\citep{Hojman1978}. This choice of torsion leads to some problems
as has been pointed out in \citep{Ni1979}. In later sections we see
that how these and similar problems can be evaded. Finally, in Section
\eqref{sec:Magnetogenesis} we see how magnetogenesis can be brought
about by assuming the restricted form.

\section{\label{sec:A-Brief-Review} Primordial Origin of Magnetic Fields}

The basic idea is that the magnetic fields are a relic of inflation.
 This  means that
the vacuum fluctuations of the electromagnetic field 
 get amplified during inflation and subsequently became classical
fluctuations in the later phases of the evolution of the universe.
This inflationary paradigm also accounts for the origin of classical
density perturbations \citep{gorbunov2011}. However, it is known
that the electromagnetic Lagrangian is conformally invariant which
is tantamount to saying that the equations of motion for the fields are
not modified by curvature induced due to a metric that is conformally
related to the Minkowski metric. Since FRW metric is conformally related
to the Minkowski metric, hence no amplification of vacuum fluctuations
can take place in this case. In fact, 
 due to spatial expansion,
 the energy density of the
electromagnetic field
varies as $a^{-4}$.
Hence, one needs to break the conformal invariance
of electromagnetism if one wants to bring about any amplification.

In \citep{Ratra1992} Ratra proposed a coupling of the form
\begin{equation}
\sqrt{-g}e^{2\alpha\phi}F^{\mu\nu}F_{\mu\nu}\label{eq:ratra_inter}
\end{equation}
between the inflaton field ($\phi)$ and electromagnetic field $F_{\mu\nu}$
with $\alpha$ as a free parameter. 
He further showed that
under slow roll inflation conditions, this causes amplification of
fields. However, it appears to pose some problems. For slow roll inflation,
let us assume that 
 $e^{\alpha\phi}$ evolves as $a^{\beta}$. 
it turns out that sufficient amount of scale invariant amplification
is achieved for some values of $\beta$ 
\citep{Ferreira2013,Tsagas:2014wba,Campanelli2015}.
However in this case either the electronic charge becomes very large
at the beginning of inflation or electric field 
 grows too fast and back reacts.
 Hence, we either end up facing
the strong coupling or back reaction problem if we demand sufficient
amplification for the magnetic field.
 A nice description of both these problems
can be found in \citep{Ferreira2013}. In an alternate treatment of this problem,
however, it is possible to 
get sufficiently strong magnetic fields with no back reaction
\citep{Tsagas:2014wba,Campanelli2015}.
In our analysis we shall follow the formalism developed in these papers.

\section{\label{sec:Minimal-coupling-and}A Brief Review of Poincar{\'e} gauge
theory}

The three fundamental interactions -- weak, 
electromagnetic and strong
are described within the framework of relativistic quantum field
theories on flat Minkowski spacetime. These quantum fields reside
in the spacetime but are aloof from it \citep{hehl1976}. Moreover
all of these theories are gauge theories. Gravity seems to be different
from these in that (a) it is a classical theory \& (b) 
 it is based on the deformation the spacetime itself. Upon
quantization, one faces the problem of non-renormalizability \citep{birrell1984,freydoon1976}. 
It is suggested that gauging, in addition to providing many other
features, might provide a renormalizable version of gravity \citep{Hehl1994}.
Besides these reasons, it is natural to inquire whether gravitation
can also be based on local gauge invariance \citep{freydoon1976,Kibble1961}.
In addition to this it has been shown in the singularity theorems
of Hawking and Penrose that a cosmology based on Riemannian geometry
would either force universe to fall into or come out of singularity.
The simplest way of avoiding such consequences is to assume non-Riemannian
geometries \citep{Hehl1994,poplaw2012}. Poincar{\'e} gauge theory based
on the local gauge invariance of the Poincar{\'e} group provides one such
paradigm.

Mathematically speaking, Poincar{\'e} group is a semi direct product\footnote{Semi direct product is a special case of direct product of two groups,
when one of the groups under consideration happens to be a normal
group. In the present case $SO\left(1,3\right)$ is the normal subgroup
of the Poincar{\'e} group.} of the spacetime translations and the Lorentz group, i.e., $P\left(1,3\right)\equiv SO\left(1,3\right)\rtimes T\left(1,3\right)$
\citep{Ali2009}. When we gauge this group, we find that instead of
one, two gauge fields are obtained. Historically, for the first time
Utiyama applied gauge principle to generate gravitational interactions
\citep{Utiyama1956} by gauging the Lorentz group. Later Kibble gauged
the whole Poincar{\'e} group \citep{Kibble1961}. Here we must emphasize
that different spacetime geometries are obtained upon gauging different
groups. For example if one considers only the translational group
then one obtains Weitzenb\"ock geometry, a geometry with torsion
but no curvature \citep{Hehl2012}. Poincar{\'e} gauge theory also assumes
the metricity condition:
\begin{equation}
\nabla_{\mu}g_{\rho\sigma}=\partial_{\mu}g_{\rho\sigma}-\Gamma_{\ \mu\rho}^{\tau}g_{\tau\sigma}-\Gamma_{\ \mu\sigma}^{\tau}g_{\rho\tau}=0,\label{eq:metricity}
\end{equation}
here the covariant derivative is taken w.r.t. the affine connection
$\Gamma_{\ \beta\gamma}^{\alpha}$. We point out 
 that there are more general classes of theories known
as metric affine theories where even metricity condition (i.e. Eq.
\ref{eq:metricity}) isn't assumed. The reader is referred to Refs.
\citep{Hehl1994,blagojevic2013} for more details.

According to general relativity, Minkowskian spacetime in the presence
of matter becomes Riemannian. The geometry of a manifold is encoded
in the connection $\Gamma_{\ \beta\gamma}^{\alpha}$ which in general
can be written as \citep{carroll2013}
\begin{equation}
\Gamma_{\ \beta\gamma}^{\alpha}=\mathring{\Gamma}_{\ \beta\gamma}^{\alpha}+K_{\ \beta\gamma}^{\alpha}\label{eq:affine_decomp}
\end{equation}
here the quantity $K_{\ \beta\gamma}^{\alpha}$ is called the contortion
and $\mathring{\Gamma}_{\ \beta\gamma}^{\alpha}=\mathring{\Gamma}_{\ \gamma\beta}^{\alpha}$
is the usual Christoffel symbol. Furthermore, the quantity
\begin{equation}
T_{\ \beta\gamma}^{\alpha}=\Gamma_{\ \beta\gamma}^{\alpha}-\Gamma_{\ \gamma\beta}^{\alpha},\label{eq:torsion_def_connec}
\end{equation}
is called the torsion, which is clearly antisymmetric w.r.t. to its
last two indices. In addition to the condition \ref{eq:metricity},
we also assume $\mathring{\nabla}_{\mu}g_{\rho\sigma}=0$. This, together
with $\mathring{\Gamma}_{\ \beta\gamma}^{\alpha}=\mathring{\Gamma}_{\ \gamma\beta}^{\alpha}$
allows us to solve for $\mathring{\Gamma}_{\ \beta\gamma}^{\alpha}$
in terms of the derivative of the metric tensor. Further more using
$\mathring{\nabla}_{\mu}g_{\rho\sigma}=0$ in Eq. \ref{eq:metricity}
gives the symmetry property of the contortion tensor
\begin{equation}
K_{\sigma\mu\rho}=-K_{\rho\mu\sigma},\label{eq:cont_anti_symm}
\end{equation}
thus the contortion tensor is antisymmetric with respect to its first
and third indices. Next, using Eq. \eqref{eq:affine_decomp} in \eqref{eq:torsion_def_connec}
we get
\begin{equation}
T_{\ \beta\gamma}^{\alpha}=K_{\ \beta\gamma}^{\alpha}-K_{\ \gamma\beta}^{\alpha}.\label{eq:cont_tors_relation}
\end{equation}
Equation \ref{eq:cont_tors_relation} using condition \ref{eq:cont_anti_symm}
can be inverted and contortion can be expressed in terms of torsion
as follows
\begin{equation}
K_{\ \beta\gamma}^{\alpha}=\frac{1}{2}\left(T_{\ \beta\gamma}^{\alpha}+T_{\beta\ \gamma}^{\ \alpha}+T_{\gamma\ \beta}^{\ \alpha}\right).\label{eq:contor_tors_rel2}
\end{equation}
In general relativity the
connection is torsion free but on account of gauging the Poincar\'e
group it becomes endowed with torsion and spacetime manifold becomes
Riemann Cartan \citep{Ali2009,Hehl2012}. So if we consider Lagrangian
proportional to Ricci scalar, which can be written as \citep{Shapiro2001}
\begin{equation}
R=\mathring{R}+\left[2\nabla_{\rho}K_{\cdot\cdot\sigma}^{\rho\sigma}+K^{\rho\alpha\nu}K_{\alpha\rho\nu}-K_{\cdot\rho\alpha}^{\rho}K_{\cdot\cdot\sigma}^{\alpha\sigma}\right],\label{eq:ricci_scalar_torsion}
\end{equation}
then $R$ gets contribution from torsion as well. The resulting theory
is called Einstein-Cartan-Sciama-Kibble (ECSK) Theory \citep{poplaw2012}.
This is a theory which approximately resembles Einstein's general
relativity but also predicts additional effects which arise due to
torsion. 

\section{Minimal Coupling and Torsion in Poincar\'e Gauge Theory}

The effects of torsion are expected to be very small in the weak field
limit. However these might play a significant role in the early Universe
\citep{Popla2012b,Hehl1974}. Introducing electromagnetism and other
gauge interactions in this framework turns out to be difficult because
it breaks gauge invariance. As a remedy, a minimal coupling procedure
has been prescribed in \citep{Hojman1978}. This model, however, is
ruled out by solar observations \citep{Ni1979}. We will see how by
coupling torsion with non-abelian gauge fields this problem can be
evaded. 
We also find it fascinating that the model \citep{Hojman1978} leads to precisely
the same form of interaction (i.e., Eq. \ref{eq:ratra_inter}) that
was proposed by Ratra.

\subsection{Torsion and Electromagnetism}

In the presence of torsion, the electromagnetic gauge covariant derivative
can be expressed as
\begin{equation}
\nabla_{\mu}A_{\nu}=\partial_{\mu}A_{\nu}-\Gamma_{\ \mu\nu}^{\alpha}A_{\alpha}=\partial_{\mu}A_{\nu}-\mathring{\Gamma}_{\ \mu\nu}^{\alpha}A_{\alpha}-K_{\enskip\mu\nu}^{\alpha}A_{\alpha}=\mathring{\nabla}_{\mu}A_{\nu}-K_{\enskip\mu\nu}^{\alpha}A_{\alpha}\,.\label{eq:gravgaugecov}
\end{equation}
Furthermore, using this, electromagnetic field tensor can be written as
\begin{equation}
F_{\mu\nu}=\nabla_{\mu}A_{\nu}-\nabla_{\nu}A_{\mu}=\partial_{\mu}A_{\nu}-\partial_{\nu}A_{\mu}-T_{\enskip\mu\nu}^{\alpha}A_{\alpha}.\label{eq:field_tensor}
\end{equation}
Due to an extra term proportional to torsion, electromagnetic gauge
invariance is not preserved. It has been suggested \citep{Hojman1978}
that if we choose a specific form for torsion, and slightly modify
the gauge transformation conditions, we can restore gauge invariance
and still have a restricted version of torsion. We impose the following
form of torsion
\begin{equation}
T_{\enskip\beta\gamma}^{\alpha}=\delta_{\gamma}^{\alpha}\partial_{\beta}\phi-\delta_{\beta}^{\alpha}\partial_{\gamma}\phi,\label{eq:tor_res_form}
\end{equation}
where $\phi$ is scalar field. Using Eq. \eqref{eq:tor_res_form}
in \eqref{eq:contor_tors_rel2}, the contortion will have the following
form
\[
K_{\ \beta\gamma}^{\alpha}=g_{\beta\gamma}\partial^{\alpha}\phi-\delta_{\beta}^{\alpha}\partial_{\gamma}\phi.
\]
 Using this expression in Eq. \eqref{eq:ricci_scalar_torsion}, the
Ricci scalar in terms of $\phi$ is found to be 
\begin{equation}
R=\mathring{R}-6\partial_{\mu}\phi\partial^{\mu}\phi+\frac{2}{\sqrt{-g}}\partial_{\rho}\left(\sqrt{-g}K_{\cdot\cdot\sigma}^{\rho\sigma}\right),\label{eq:fullRicci}
\end{equation}
where $\mathring{R}$ is the standard Ricci scalar computed using
the Christoffel connection $\mathring{\Gamma}_{\ \beta\gamma}^{\alpha}$.
The gauge transformation gets modified to 
\begin{equation}
A_{\mu}\rightarrow A_{\mu}+e^{\phi}\partial_{\mu}\epsilon,
\label{eq:gaugetransA}
\end{equation}
where $\epsilon$ is the transformation parameter. We see that we
obtain an extra contribution equal to $e^{\phi}$ which vanishes in
the absence of torsion. In this case, it can be easily checked that
the field tensor, Eq. \eqref{eq:field_tensor}, and hence the Lagrangian
remains invariant. 

 The modified form of gauge transformation can
now be extended to charged matter fields. Let $\psi$ denote a complex
scalar field which transforms under U(1) gauge transformation as
$\psi\rightarrow \psi' = \exp(ie\epsilon)\psi$. The minimal coupling
prescription \cite{Hojman1978} leads to the covariant derivative $D_\mu\psi = 
\partial_\mu\psi - ie\exp(-\phi)A_\mu\psi$.  
Hence we find that the effective electromagnetic coupling in this case
is $e/\exp(\phi)$.
Using the modified form
of torsion in the gravitational Lagrangian, we obtain 
\begin{equation}
S=\int d^{4}x\sqrt{-g}\left[-\frac{M_{\text{pl}}^{2}}{16\pi}\left(\mathring{R}-6\partial^{\rho}\phi\partial_{\rho}\phi\right)-\frac{1}{4}F^{\mu\nu}F_{\mu\nu}
+(D_\mu\psi)^* D^\mu\psi  \right],
\label{eq:action1}
\end{equation}
where the effects of torsion have been explicitly displayed in terms
of the field $\phi$ and $M_{\text{pl}}$ is the Planck mass. Here
we have performed an integration by parts and dropped a total divergence.
In order to relate this to the form given in 
 in Eq. \eqref{eq:ratra_inter}, we perform the
transformation, 
\begin{equation}
A^{\mu}=V^{\mu}e^{\phi}\label{eq:scalingvecpot}
\end{equation}
In terms of the field $V_{\mu}$, the gauge transformation of Eq.
\eqref{eq:gaugetransA} becomes $V_{\mu}\rightarrow V_{\mu}+\partial_{\mu}\epsilon$
and the field tensor becomes, 
\begin{equation}
F_{\mu\nu}(A)=e^{\phi}F_{\mu\nu}(V).
\end{equation}
 The covariant derivative $D_\mu\psi = 
\partial_\mu\psi - ieA_\mu\psi$ now takes the standard form and  
the action can be written as 
\begin{equation}
S=\int d^{4}x\sqrt{-g}\left[-\frac{M_{\text{pl}}^{2}}{16\pi}\left(\mathring{R}-6\partial^{\rho}\phi\partial_{\rho}\phi\right)-\frac{1}{4}e^{2\phi}F^{\mu\nu}F_{\mu\nu} +
(D_\mu\psi)^* D^\mu\psi
\right]\label{eq:minimalcoupling}
\end{equation}
Now, we can see the similarity with the Lagrangian in Ratra model \cite{Ratra1992}
during inflation. It is clear that the coupling term $f^{2}$ which
had to be put by hand in the Ratra model has automatically come out.
  We point out that this term could be inserted in the coupling to matter
fields or directly as a coupling with the gauge kinetic term in the 
Ratra model also exactly in analogy with Eqs. \ref{eq:action1}
and \ref{eq:minimalcoupling}. 

However the Hojman et al minimal prescription 
model \cite{Hojman1978} (Eq. \ref{eq:minimalcoupling}) 
is in conflict with solar data \citep{Ni1979}.
As we shall see in the next two subsections this problem is evaded
since the minimal coupling procedure extended to non-abelian gauge
fields naturally generates effective potential terms for the field
$\phi$. Alternatively we may go beyond the minimal coupling prescription
\citep{Hojman1978} by adding kinetic and potential terms in the field
$\phi$ while preserving Poincar{\'e} gauge invariance.

Before we end this subsection we briefly discuss an alternative
scenario that has been proposed for handling internal gauge symmetries
in the case of Poincar{\'e} theory. It has been argued in the literature that 
gauge fields associated with internal symmetries, such as the photon field,
 should
not be coupled to torsion \cite{hehl1976,Blagojevic:2003cg}. This is another way to avoid the problem of
violation of gauge invariance associated with internal groups. 
However a massive spin 1 field 
must necessarily couple to torsion. This presents a paradox since a
massive spin 1 field can also be generated from an internal gauge symmetry
by the Higgs mechanism. For example, 
lets consider a U(1) gauge field coupled to a charged scalar particle. Let
us use the
standard version of the gauge covariant derivative, 
$ \nabla_\mu A_\nu = \partial_\mu A_\nu $,
i.e. we do not introduce the contortion tensor in this definition following
the prescription given in \cite{hehl1976}. In contrast
if we considered a massive spin 1 field we will need to introduce the
contortion term \cite{Blagojevic:2003cg}. 

Now lets consider the coupling of the U(1) gauge field with a
 charged scalar. The Lagrangian is
\begin{equation}
{\cal L} = {1\over 2} g^{\mu\nu} (D_\mu\phi)^* D_\nu\phi -V
\end{equation}
where $ D_\mu\phi = (\partial_\mu-ieA_\mu)\phi$ is the U(1) gauge covariant
derivative. We assume Higgs mechanism, i.e. the potential is such that
vacuum value of $\phi$ is not zero,
$\langle \phi\rangle =v$.
In this case we find that the gauge field $A_\mu$ becomes massive. In the
unitary gauge this field behaves identically to a massive vector
field. But the problem is that it does not couple to the torsion tensor.
Hence it does not behave as the
usual massive spin 1 field. This is clearly inconsistent since a massive
vector field must necessarily couple to torsion. In this case we will
have to further generalize our prescription in order to forbid some class
of massive spin-1 fields from coupling to torsion. Hence we argue that it 
is better to explore the framework presented in \citep{Hojman1978} since
it does not lead to such inconsistencies.

\subsection{\label{sec:nonabelian}Torsion and Non-abelian Fields}

In this section we generalize the principle of minimal coupling \citep{Hojman1978}
to make torsion compatible with non-abelian fields as well. Consider
a matter field $\psi$ which under a non abelian group $G$ transforms
as 
\begin{equation}
\psi^{\prime}(x)=U(x)\psi(x)
\end{equation}
where $U\in G$ is a non-abelian transformation given by 
\begin{equation}
U(x)=\exp[i\alpha^{i}(x)T^{i}],
\end{equation}
$\alpha^{i}$'s are transformation parameters and $T^{i}$'s are generators
of the group $G$. 
In analogy with 
Eq. \ref{eq:gaugetransA} we propose the following transformation for
the gauge fields $W^{i}$, 
\begin{equation}
W_{\mu}^{i\prime}T^{i}=U\left(W_{\mu}^{i}T^{i}+e^{\phi}\frac{i}{g}\partial_{\mu}\right)U^{\dagger},\label{eq:gaugetransW}
\end{equation}
with the gauge derivative given by 
\begin{equation}
D_{\mu}\psi=\left(\partial_{\mu}-ige^{-\phi}W_{\mu}^{i}T^{i}\right)\psi.
\end{equation}
The covariant derivative transforms as 
\begin{equation}
\left(D_{\mu}\psi\right)^{\prime}=U(x)D_{\mu}\psi.
\end{equation}
Furthermore, we obtain the field strength tensor by computing $[D_{\mu},D_{\nu}]$.
This leads to 
\begin{equation}
[D_{\mu},D_{\nu}]\psi=-ige^{-\phi}F_{\mu\nu}^{i}T^{i}\psi,
\end{equation}
where 
\begin{equation}
F_{\mu\nu}^{i}T^{i}=\partial_{\mu}W_{\nu}^{i}T^{i}-\partial_{\nu}W_{\mu}^{i}T^{i}-ige^{-\phi}[W_{\mu}^{i}T^{i},W_{\nu}^{j}T^{j}]-W_{\nu}^{i}T^{i}\partial_{\mu}\phi+W_{\mu}^{i}T^{i}\partial_{\nu}\phi.
\end{equation}
One can easily check that this is just the expression obtained by
replacing the derivatives by gravitational gauge covariant derivatives
(see Eq. \ref{eq:gravgaugecov}), i.e., 
\begin{equation}
F_{\mu\nu}^{i}T^{i}=\nabla_{\mu}W_{\nu}^{i}T^{i}-\nabla_{\nu}W_{\mu}^{i}T^{i}-ige^{-\phi}[W_{\mu}^{i}T^{i},W_{\nu}^{j}T^{j}]
\end{equation}
We can now make the transformation of the non-abelian vector potential
$W_{\mu}^{i}$ analogous to Eq. \ref{eq:scalingvecpot}. This leads
to the standard form of the field strength tensor up to an overall
factor of $e^{\phi}$. Hence the kinetic energy term of the non-abelian
fields becomes $-(1/4)e^{2\phi}F_{\mu\nu}^{i}F^{i\mu\nu}$ as in the
case of abelian gauge theory.

The coupling of $\phi$ with non-abelian gauge fields is interesting
since it generates an effective potential for $\phi$. Consider a
$SU(3)$ gauge field analogous to QCD. After dynamical symmetry breaking,
the operator $F_{\mu\nu}F^{\mu\nu}$ acquires an expectation value,
i.e. 
\begin{equation}
\langle F_{\mu\nu}F^{\mu\nu}\rangle=\Lambda^{4}
\end{equation}
where $\Lambda$ is a parameter. At leading order we may therefore
replace the term in the Lagrangian as 
\begin{equation}
e^{2\phi}F_{\mu\nu}F^{\mu\nu}\rightarrow e^{2\phi}\langle F_{\mu\nu}F^{\mu\nu}\rangle\label{eq:effectiveVQCD}
\end{equation}
which leads to an effective potential term for $\phi$. We will also
get additional contribution from the topological term $\epsilon^{\mu\nu\alpha\beta}F_{\mu\nu}F_{\alpha\beta}$
which will also pick up an overall factor $\exp(2\phi)$. We point
out that other potential terms also get generated, the details of
which depend on the model being considered. We shall discuss an explicit
model below.

The potential terms are expected to lead to a background value $\phi_{0}$
of the field $\phi$. As the Universe evolves, we assume that $\phi$
also undergoes a slow cosmological evolution and takes a value $\phi_{0}$
at the current time. The evolution should be sufficiently slow in
order that it is not in conflict with constraints on time dependence
of fundamental parameters. We next expand $\phi$ about its background
value such that 
\begin{equation}
\phi=\phi_{0}+\hat{\phi}.
\end{equation}
Furthermore we scale $\phi$ in order to convert its kinetic energy
term into the canonical form. The background factor $e^{\phi_{0}}$
gets absorbed into redefinition of parameters. With this expansion
we obtain a mass term for $\hat{\phi}$ as well as higher order terms.
This field acquires an effective mass given by 
\begin{equation}
m_{\phi}\sim\frac{\Lambda^{2}}{\beta M_{\text{pl}}}\label{eq:scalar_field_mass}
\end{equation}
Hence we generate a mass term as well as higher order potential terms
for the field $\phi$. We discuss this in more detail below.

\subsection{Evading the Solar Constraints}

As has been discussed in literature \citep{Ni1979}, the minimal coupling
model discussed above is in conflict with constraints from the solar
system. It turns out that the torsion or the scalar field $\phi$
generated by the Sun is sufficiently large that it leads to observable
differences between the gravitational accelerations of particles with
different electromagnetic energy content. Non-observation of this
difference rules out the minimal coupling model \citep{Hojman1978}
discussed above. However the constraints can be evaded if we add a
suitable potential $V(\phi)$ to the action in Eq. \eqref{eq:actiontorsion}
such that the scalar field acquires mass. As we have shown in the
previous section, we may not need to explicitly add potential terms
since an effective potential also gets generated by non-abelian gauge
fields. Once the field acquires mass, its equation of motion can be
expressed as 
\begin{equation}
\nabla^{2}\phi-m_{\phi}^{2}\phi=\frac{1}{3}\left(\mathbf{B}^{2}-\mathbf{E}^{2}\right).
\end{equation}
This is a generalization of Eq. 29 of \citep{Ni1979}. The electromagnetic
energy content of the Sun acts as a source of the field $\phi$. We
see that if the mass $m_{\phi}$ is sufficiently large then the field
$\phi$ will decay exponentially and will be negligible at Earth.
Assuming $\beta$ in Eq. \eqref{eq:scalar_field_mass} of order unity
and $\Lambda$ of order 1 GeV corresponding to the QCD scale, we obtain
$m_{\phi}$ of order $10^{-10}$ eV. This is large enough to completely
suppress the signal arising due to Sun.

We point out that besides the potential term generated through QCD
(Eq. \ref{eq:effectiveVQCD}), quantum corrections due to electroweak,
gravity and other beyond the Standard Model fields would also generate
other terms in the effective potential for the $\phi$ field. Hence
it is natural to include potential terms for this field.

\section{Generalized Gravitational Action}

The minimal coupling model implied by the Poincar{\'e} symmetry has the
necessary ingredients to generate magnetogenesis. It naturally produces
the coupling of a scalar field $\phi$ with the electromagnetic field
similar to that assumed in \citep{Ratra1992}. As we have discussed
in Section \eqref{sec:nonabelian} that a potential for the field
$\phi$ also gets generated naturally.  
Here we shall not necessarily assume that $\phi$ is also the inflaton
field. It is primarily responsible for magnetogenesis.
 In the next section, we shall study the
generation of magnetic fields within the minimal model. As we shall
see, the minimal model Eq. \eqref{eq:minimalcoupling} can lead to
an enhancement of the magnetic fields. In this section,
we study generalized models which may lead to a modification of the
kinetic energy term of the field $\phi$. This may be useful in avoiding
the back reaction problems. 

Before discussing the generalized models, we point out that it is
possible to add kinetic and potential terms in the field $\phi$ while
preserving Poincar{\'e} and the electromagnetic gauge invariance which
are the guiding principles in constructing this action. In particular
we can add terms in gravitational action including the field $\phi$,
such as, 
\begin{equation}
L_{T}=\tilde{\beta}g^{\gamma\sigma}T_{\ \beta\gamma}^{\alpha}T_{\ \alpha\sigma}^{\beta}.\label{eq:torsionKE}
\end{equation}
This directly leads to a kinetic term for $\phi$. In contrast we
cannot generate a potential term for $\phi$ by adding terms involving
the torsion tensor. Such terms have to be added directly in terms
of $\phi$. However as we have seen, an effective potential is generated
by non-abelian QCD like fields. After adding such terms, the final
action can be expressed as:
\begin{equation}
S=\int d^{4}x\sqrt{-g}\left[-\frac{M_{\text{pl}}^{2}}{16\pi}\left(\mathring{R}-6\beta^{2}\partial^{\rho}\phi\partial_{\rho}\phi\right)-\frac{1}{4}e^{2\phi}F^{\mu\nu}F_{\mu\nu}-V(\phi)\right]
\label{eq:actiontorsion}
\end{equation}
where the added kinetic term has been accommodated by introducing
the parameter $\beta$.

Following Campanelli \citep{Campanelli2015}, we find that this action
can generate magnetic field of the required strength. Depending on
the model, i.e. the choice of potential, it may be necessary to choose
$\beta$ to be very small. This will require fine tuning since the
additional kinetic energy term has to be chosen very precisely in order
to obtain a small value of $\beta$. This fine tuning may be evaded
if we further generalize the gravitational action such that it becomes,
\begin{equation}
S_{g}=\int d^{4}x\sqrt{-g}\left[-e^{2\phi}\frac{M_{\text{pl}}^{2}}{16\pi}R-
\frac{1}{6}e^{2\phi}M_{T}^{2}g^{\mu\nu}T_{\ \mu\beta}^{\alpha}T_{\ \alpha\nu}^{\beta}-V(\phi)\right]\label{eq:actiontorsion1}
\end{equation}
Here the full form of $R$ is given in Eq. \eqref{eq:fullRicci}.
We next make a conformal transformation such that 
\begin{equation}
\tilde{g}_{\mu\nu}=e^{2\phi}g_{\mu\nu}\label{eq:conformaltrans}
\end{equation}
In terms of the new metric $\tilde{g}_{\mu\nu}$, the gravitational
action becomes 
\begin{equation}
S_{g}=\int d^{4}x\sqrt{-\tilde{g}}\left[-\frac{M_{\text{pl}}^{2}}{16\pi}\mathring{R}+\frac{1}{2}M_{T}^{2}\tilde{g}^{\mu\nu}\partial_{\mu}\phi\partial_{\nu}\phi-V(\phi)\right]\label{eq:actiontorsion2}
\end{equation}
Here we do not need to fine tune the parameter $M_{T}$. 
Now we may choose the potential
such that $\phi$ may act as the inflaton field. The matter action
is chosen by the principle of minimal coupling as prescribed in \citep{Hojman1978}.
The full action has precisely the form which leads to both inflation
and magnetogenesis following the analysis presented in \citep{Ratra1992,Campanelli2015}.

Due to the conformal transformation Eq. \eqref{eq:conformaltrans}
the matter action also undergoes some change. The gauge kinetic energy
terms remain unaffected. However terms involving scalar and spinor
fields may change. Lets us consider the effect on the Higgs field
$H$. With the transformation Eq. \eqref{eq:conformaltrans}, the
Higgs action becomes 
\begin{equation}
S_{H}=\int d^{4}x\sqrt{-\tilde{g}}\left[e^{-2\phi}\tilde{g}^{\mu\nu}D_{\mu}H^{\dagger}D_{\nu}H-m_{H}^{2}e^{-4\phi}H^{\dagger}H-\lambda e^{-4\phi}(H^{\dagger}H)^{2}\right]
\end{equation}
where $D_{\mu}$ is the gauge covariant derivative. Similar terms
are generated for all scalar fields which have non-zero vacuum value.
Furthermore the fermion field condensates are expected to generate
additional terms in the effective potential for $\phi$. In order
to reduce Higgs kinetic energy term to its canonical form, we may
transform the Higgs field such that 
\begin{equation}
\tilde{H}=e^{-\phi}H
\end{equation}
This transformation, however, leads to complicated derivative terms
for the $\phi$ field. In any case, we find the appearance of additional
potential terms for $\phi$ from the scalar field action besides the
terms discussed in Section \eqref{sec:nonabelian}.

Working with the field $H$, i.e. without transforming to $\tilde{H}$,
we may express the potential terms as 
\begin{equation}
V(\phi)=ae^{-4\phi}+be^{2\phi}+...
\end{equation}
The factors $a$ and $b$ represent the contributions due to the vacuum
expectation value of the Higgs field and the QCD condensates. These
are not really constant since these will also change as $\phi$ evolves
with time. In order to determine the vacuum expectation value of $\phi$
we can treat $a$ and $b$ as constants. Keeping only these two terms
the minimum of the potential is found to be 
\begin{equation}
\phi_{0}=\frac{1}{6}\ln(2a/b)
\end{equation}
By expanding around the minimum and rescaling $\phi$ such that $\tilde{\phi}=M_{T}\phi$
we obtain a mass term for $\tilde{\phi}$. The overall terms, i.e.
$\exp(-4\phi_{0})$ and $\exp(2\phi_{0})$, should be absorbed into
the Higgs vacuum expectation value and the QCD condensates respectively.
The $\phi$ mass is found to be order
\[
\frac{1}{\sqrt{M_{T}}}\max\big\{ \sqrt{a},\sqrt{b}\big\} .
\]
We expect $a$ to be at least as large as the electroweak scale and
$M_{T}<M_{\text{pl}}$. This generates a mass larger than $10^{-6}$
eV which is sufficient to evade the solar constraints.

For generation of primordial magnetic fields, however, the situation
is more complicated. This is because now we need to study the time
evolution of the Higgs as well as the QCD field. In case of QCD this
will require a rather complicated quantum analysis of non-abelian
fields. We do not pursue this in the present paper and simply assume
a potential for $\phi$ which can lead to primordial magnetogenesis.
As discussed earlier, quantum corrections would generate additional
terms in the effective potential for $\phi$ which need to be included
for a complete analysis.

The introduction of the factor $e^{2\phi}$ in Eq. \ref{eq:actiontorsion1}
is partially justified by considering an $f(R)$ gravity model such
as the Starobinsky model \citep{Starobinsky1980}. 
 In that case the gravitational action becomes
very complicated due to the presence of the $R^{2}$ term which involves
four derivative terms of the field $\phi$. However if we generalize
it such that the action reads, 
\begin{equation}
S_{\text{Staro}}=\int d^{4}x\sqrt{-g}\left[-\frac{M_{\text{pl}}^{2}}{16\pi}\left(e^{2\phi}R-\frac{R^{2}}{6M^{2}}\right)+e^{2\phi}M_{T}^{2}g^{\mu\nu}T_{\ \mu\beta}^{\alpha}T_{\ \alpha\nu}^{\beta}\right]\label{eq:actionStaro}
\end{equation}
then after the conformal transformation the action reduces to the
standard Starobinsky action along with an additional kinetic energy
term for the field $\phi$ given in Eq. \ref{eq:actiontorsion}. Only
with the introduction of this extra factor $e^{2\phi}$ do we get
a simple action in the Einstein frame.

An alternative approach is to demand global conformal invariance.
In this case the mass scale $M_{\text{pl}}$ gets replaced by $a\Phi$
where $\Phi$ is a scalar field and $a$ is a parameter, see for example
\citep{Jain2011}. We do not introduce the extra kinetic energy term
proportional to $M_{T}$. The resulting gravitational action may be
expressed as, 
\begin{equation}
S_{g}=\int d^{4}x\sqrt{-g}\left[\frac{(a\Phi)^{2}}{16\pi}R+\frac{1}{2}g^{\mu\nu}\partial_{\mu}\Phi\partial_{\nu}\Phi-V(\Phi,\phi)\right]\label{eq:actionconformal}
\end{equation}
In this case we do not need to introduce the extra factor $e^{2\phi}$
in the gravitational action. This action is invariant under the global
conformal transformation $g_{\mu\nu}\rightarrow g_{\mu\nu}/\Lambda^{2}$,
$\Phi\rightarrow\Lambda\Phi$. We impose this symmetry also on the
matter action. We next make the transformation $\tilde{g}_{\mu\nu}=e^{2\phi}g_{\mu\nu}$,
$\tilde{\Phi}=e^{-\phi}\Phi$. The action now becomes 
\begin{equation}
S_{g}=\int d^{4}x\sqrt{-\tilde{g}}\left[\frac{(a\tilde{\Phi})^{2}}{16\pi}\mathring{R}+{\cal L}_{K}-V(\tilde{\Phi},\phi)\right]\label{eq:actionconformal1}
\end{equation}
where 
\begin{equation}
{\cal L}_{K}=\frac{1}{2}\tilde{g}^{\mu\nu}\left(\partial_{\mu}\tilde{\Phi}\partial_{\nu}\tilde{\Phi}+\tilde{\Phi}^{2}\partial_{\mu}\phi\partial_{\nu}\phi-2\tilde{\Phi}\partial_{\mu}\tilde{\Phi}\partial_{\nu}\phi\right)
\label{eq:actionconformal2}
\end{equation}
The conformal symmetry may be broken softly by the dynamical mechanism
analogous to the one described in \citep{Hur2011,Meissner2006,Meissner2007,Jain2015,Jain2014}.
Here one assumes the existence of a dark strongly coupled sector.
The condensate formation in this sector leads to dynamical breakdown
of conformal invariance which also triggers the electroweak spontaneous
symmetry breaking. Here we assume some strongly coupled sector with
very high mass scale such that these particles decay at some early
time during the evolution of the Universe. The condensate formation
in this sector leads also to a vacuum expectation value of the field
$\tilde{\Phi}$. We assume that this field undergoes negligible evolution
since the beginning of inflation and hence we can simply set it equal
to its vacuum expectation value. Alternatively we need to go to the
Einstein frame. In the present case the additional terms generated
by going from Jordan to Einstein frame are assumed to be negligible
at leading order due to our assumption that $\tilde{\Phi}$ evolves
negligibly during inflation. We therefore set $a\tilde{\Phi}$ equal
to its vacuum expectation value which is assumed to be equal to $M_{\text{pl}}$.
Ignoring the derivatives of $\tilde{\Phi}$, we generate a kinetic
energy term for $\phi$ whose overall normalization is equal to $\tilde{\Phi}^{2}$.
We assume that this value is sufficiently small in comparison to $M_{\mathrm{pl}}$
so that it does not lead to back reaction during inflation. This requires
us to set the parameter $a$ of the order of $10^{3}$. Hence with
an appropriate choice of the parameter $a$ this leads to exactly
the action given in Eq. \ref{eq:actiontorsion2} with the scale $M_{T}$
being generated by $\tilde{\Phi}$.

Before ending this section we propose a generalization of the Starobinsky
model by demanding conformal symmetry. The resulting action can be
written as 
\begin{equation}
S_{C}=\int d^{4}x\sqrt{-g}\left[-\frac{1}{16\pi}\left(a^{2}\Phi^{2}R-\xi^{2}R^{2}\right)+\frac{1}{2}g^{\mu\nu}\partial_{\mu}\Phi\partial_{\nu}\Phi-V(\Phi,\phi)\right]\label{eq:actionStaro1}
\end{equation}
We again make the transformation $\tilde{g}_{\mu\nu}=e^{2\phi}g_{\mu\nu}$,
$\tilde{\Phi}=e^{-\phi}\Phi$. The resulting action reads 
\begin{equation}
S_{C}=\int d^{4}x\sqrt{-\tilde{g}}\left[-\frac{1}{16\pi}\left(a^{2}\tilde{\Phi}^{2}\mathring{R}-\xi^{2}\mathring{R}^{2}\right)+{\cal L}_{K}-V(\tilde{\Phi},\phi)\right]\label{eq:actionStaro2}
\end{equation}
As discussed earlier the field $\tilde{\Phi}$ acquires vacuum expectation
value such that $a\tilde{\Phi}=M_{p}$. Assuming that $\tilde{\Phi}$
does not evolve significantly with time during inflation we can replace
it with its vacuum expectation value. Then the first two terms on
the right hand side yield the standard Starobinsky model. The remaining
terms involve the kinetic and potential terms for the field $\phi$
and $\tilde{\Phi}$. Along with the coupling with the electromagnetic
field these can lead to magnetogenesis.

To summarize this section, we have shown that the model for magnetogenesis
which was proposed by Ratra \citep{Ratra1992} naturally appears in
the Poincar\'e invariant theory with the minimal prescription principle
proposed in \citep{Hojman1978}. The minimal model, however, requires
fine tuning of the kinetic energy term and does not contain a potential
term for scalar field $\phi$ which generates torsion. We have argued
that potential terms for this field appear naturally and it is easy
to generalize the model such that it does not suffer from fine tuning.
This is best accomplished by imposing conformal invariance. Hence
the model has all the ingredients required for primordial magnetogenesis.
We have also described a conformal generalization of the Starobinsky
model which will lead to both inflation and magnetogenesis.

\section{\label{sec:Magnetogenesis}Magnetogenesis}

The basic formalism for magnetogenesis in the Ratra model has been
developed in several papers \cite{Ferreira2013,Campanelli2015}. Our basic point is
that Poincar\'e gauge theory naturally leads to models similar to the
Ratra model. In particular the minimal coupling model leads to a specific
form of the coupling of a scalar field $\phi$ to the electromagnetic
fields as well as the kinetic term for $\phi$. As we have discussed
in the previous section, both the kinetic and potential terms in $\phi$
may be substantially modified in comparison to what we obtain by the
minimal coupling procedure. Hence depending on the model, different
scenarios described in \citep{Campanelli2015} could be realized.
Hence it is clear that we can always generate magnetic field of the
required strength if we go beyond the minimal coupling model. In this
section we discuss how far the minimal coupling model is successful
in generating magnetic fields.

The minimal coupling model, proposed in \citep{Hojman1978}, is given
in Eq. \ref{eq:minimalcoupling}. Here we have only displayed the
coupling of $\phi$ with the electromagnetic field. Similar terms
also should be included for all gauge fields, abelian and non-abelian.
As described above, an effective potential term for the scalar field
$\phi$ of the form 
\begin{equation}
V=M_{1}^{4}e^{2\phi}\label{eq:effectivepotential}
\end{equation}
naturally appears due to the vacuum condensates formed by QCD like
non-abelian fields. Here we assume existence of some QCD like fields
with very high mass scale $M_{1}$. The equation of motion for $\phi$,
neglecting electromagnetic effects, can be expressed as 
\begin{equation}
12M_{\text{pl}}^{2}\beta^{2}(\ddot{\phi}+3H\dot{\phi})+\frac{\partial V}{\partial\phi}=0
\label{eq:eqofmotuon}
\end{equation}
with $\beta=1$. Here we assume that inflation is caused by some field
other than $\phi$. Let us assume that during inflation the time dependent
part of this field is small. In that case we can approximate $\exp(2\phi)\approx1+2\phi$.
This leads to a linear potential $V(\phi)\approx b\phi$ for $\phi$
and admits a solution of the form 
\begin{equation}
e^{\phi}=C\left(\frac{a}{a_{i}}\right)^{\alpha}
\label{eq:defnalpha}
\end{equation}
with constant $\alpha\ll1$. Ignoring the constant part of $\phi$,
we obtain $\phi\approx\alpha Ht$. The solution starts to break down
as $\phi$ approaches unity. This corresponds to the small $p$ ($p$ is same
as $\alpha$ in our notation) limit
of the models discussed in \citep{Campanelli2015}. In this limit
the model leads to magnetic fields of order $10^{-32}$ G which is rather small.

We next discuss the more general case in which the field $\phi$ is
not necessarily small while working within the framework of the minimal
coupling model. In this case we use the full form of the potential
given in Eq. \ref{eq:effectivepotential}. Assuming that $\ddot{\phi}$
is negligible, the solution for $\phi$ becomes 
\begin{equation}
e^{2\phi}=\frac{1}{2m(t+t_{0})}
\end{equation}
where $m=M_{1}^{4}/(18M_{\text{pl}}^{2}H)$ and $t_{0}$ is an integration
constant. We set $t_{0}=N_{0}/H$ and find that $\ddot{\phi}$ is
negligible if 
\begin{equation}
Ht+N_{0}\gg1
\end{equation}
which holds for a wide range of choices of $N_{0}$. We point out
that this requires $N_{0}$ to be sufficiently larger than unity.
Furthermore this condition holds independent of the value of $\beta$.
With this condition we also find that the kinetic energy term for
$\phi$ does not cause back reaction for the inflationary potential.
The effective value of $\alpha$ in this case is found to be 
\begin{equation}
\alpha=-\frac{1}{2(Ht+N_{0})}
\end{equation}
which varies slowly with time but is necessarily small throughout
inflation. Hence we find that with this choice of potential we are
forced to have small values of $\alpha$ which effectively also imply
a small value of the time dependent part of $\phi$. 
This implies
that the minimal coupling model does not lead to sufficiently large magnetic
fields which can provide seeds for the galactic dynamo. 
This can be modified
only if we allow a generalized potential which is permissible in our
framework but goes beyond the minimal coupling model. 

So far we have assumed that $\phi$ is not the inflaton field. However
since it varies slowly, the effective potential remains approximately
constant during most of the evolution. Hence it is possible to consider
$\phi$ also as the inflaton field, as long as we choose $N_{0}$
to be sufficiently large. However it is not clear how to exit inflation
and enter the reheating phase in this framework. It may be possible
to enhance the model in order to accomplish this. For example, we
may also add the term $F^{\mu\nu}\tilde{F}_{\mu\nu}$ for the non-abelian
fields which also acquires a vacuum expectation value. This term has
an effective coupling to an axion like field $\chi$ and would also
pick up a factor $e^{2\phi}$. Let us assume that this condensate
dominates the evolution during inflation. Furthermore the axion field
$\chi$ remains constant during inflation but undergoes evolution
towards the end which effectively ends inflation and leads to reheating.

\subsection{Beyond the Minimal Model}
 We have already seen that within the framework of the minimal model
we are unable to generate the magnetic field of the required strength. 
As discussed earlier, the required kinetic energy term in $\phi$ can be
 generated
naturally within the minimal model, as demonstrated, for example,
 in Eqs. \ref{eq:actionconformal} and \ref{eq:actionconformal1}. 
However we are unable to generate the required potential term.   
Here we go beyond the minimal model by adding the following potential term: 
\begin{equation}
V(\phi)=6\beta^{2}M_{p}^{2}m^{2}\phi^{2}
\end{equation}
where $m$ is a parameter with dimensions of mass. 
The kinetic energy term in $\phi$ is given in Eq. \ref{eq:actiontorsion}
or equivalently in Eq. \ref{eq:actionconformal2} with $\langle\tilde\Phi
\rangle = \sqrt{3/4\pi} \beta M_{pl}$. In this case $\phi$ also acts as
the inflaton. 

Imposing the slow roll condition
$\ddot{\phi}\ll3H\dot{\phi}$ in the equation of motion   
Eq. \ref{eq:eqofmotuon} we obtain the solution
\begin{equation} 
\phi = \phi_0 e^{-t/\tau}
\end{equation} 
where $\phi_0$ and $\tau$ are constants, such that 
\begin{equation} 
\tau=\frac{3H}{m^{2}}
\end{equation} 
The slow roll condition implies,
\begin{equation} 
\tau\gg\frac{1}{H}
\end{equation} 
which also leads to
\begin{equation} 
\frac{m^{2}}{H^{2}}\ll 1
\end{equation} 
From Eq. \ref{eq:defnalpha}, we obtain 
\begin{equation} 
\alpha = -{\phi_0\over \tau H} e^{-t/\tau} 
\end{equation} 
Now, let us estimate the value of $\phi_{0}$. The Einsteins equations
of motion give us
\begin{equation} 
3H^{2}=\frac{3}{2}\beta^{2}\dot{\phi}^{2}+3\beta^{2}m^{2}\phi^{2}
\end{equation} 
Here, for kinetic term to be negligible, we need
\begin{equation} 
\tau\gg\frac{1}{m}
\end{equation} 
Ignoring the kinetic energy term, we obtain, $\phi_0\approx H/(\beta m)$. 

Let us now assume that $\alpha\approx -2$ and approximately constant over 
much of inflation. 
The value $\alpha = -2$ is required in order to generate scale invariant magnetic field within the framework
developed in \cite{Campanelli2015}.  
This can be accomplished by requiring that $t\ll\tau$ during
inflation. This is equivalent to requiring that $H\tau \gg N$, where $N$ is
the number of e-folds during inflation. For example if we take $H\tau\approx
 500$,
assuming that $N\approx 60$, it will assure that the 
necessary conditions are met over much of inflation and we would produce a 
nearly scale invariant spectrum for a wide range of values of $k$ of the
required strength 
\citep{Campanelli2015}. The parameter values for this case would be
$m\approx H\sqrt{3/500}$ and 
 $\beta\approx 1/(-\alpha \tau m)\approx 0.5/\sqrt{1500} \ll1$.    
Hence with this choice of parameters we are able to generate the required
magnetic field.

\section{Conclusions}

The problem of generating magnetic fields during inflation can be
resolved by breaking the conformal invariance of the electromagnetic
Lagrangian \cite{Ratra1992}. The required coupling between
Torsion and Electromagnetism however had to be put in by hand. We have
shown that torsion naturally leads to this coupling within
the framework of the model developed by Hojman et al
\citep{Hojman1978} which satisfies electromagnetic gauge invariance. 
The model is based on a minimal version of torsion such that
the torsion field can be expressed in terms of a scalar field $\phi$ called
tlaplon in the literature \citep{Hojman1978}.
The main problem with this model is that it is ruled out
by constraints due to solar data \citep{Ni1979}. We have shown that
due to coupling with non-abelian fields, the minimal model acquires an 
effective potential term for the scalar field and evades the solar 
constraints. We have studied the generation of magnetic fields within
the framework of the minimal model. We find that these
are equivalent to the small $p$ (or $\alpha$) limit of the models discussed in 
\citep{Campanelli2015}. Hence the magnetic field generated in this
case is relatively small. The minimal model also leads to a rather
large contribution from the kinetic energy of $\phi$ unless $p$ is 
close to $0$.  

We have also discussed several generalizations of the minimal model.
We have argued that quantum corrections will generate additional potential
terms and hence there is no reason to restrict oneself to the minimal 
model. We can add potential terms in the scalar tlaplon while maintaining
Poincare symmetry. With these terms it is possible to generate larger
magnetic fields in comparison to the minimal model. 
In particular we have considered a model which displays invariance
under global conformal transformation  
$g_{\mu\nu}\rightarrow g_{\mu\nu}/\Lambda^{2}$,
$\Phi\rightarrow\Lambda\Phi$, where $\Phi$ is a scalar field.  
By imposing this symmetry on a generalized Starobinsky model we find
that we can naturally suppress the kinetic energy terms of $\phi$. With
this suppression it possible to have significant variation of $\phi$ during
inflation which is required for magnetogenesis.   
 By assuming a simple form of the potential for $\phi$ we have 
explicitly demonstrated the generation of nearly scale invariant magnetic
field of required strength.
However we still need a mechanism to generate the required potential which has so
far simply been assumed. For this purpose it may be useful to study the conformal
model in more detail using the effective potential approach which includes corrections
due to loop contributions. 

\bigskip
\noindent
{\bf Acknowledgements} We thank Bharat Ratra for useful comments on the paper.

\bibliography{magnetic}

\end{document}